\documentclass[a4paper,12pt]{article}
\usepackage{graphics}
\usepackage{amsmath,amssymb}
\usepackage{color}
\def\vf{\varphi}

\def\half{\frac{1}{2}}

\def\be{\begin{equation}}
\def\ee{\end{equation}}
\def\bea{\begin{eqnarray}}
\def\eea{\end{eqnarray}}
\def\beax{\begin{eqnarray*}}
\def\eeax{\end{eqnarray*}}
\begin{document}
\setcounter{footnote}{1}
\renewcommand{\thefootnote}{{\fnsymbol{footnote}}}
\title{Which Green Functions Does the Path Integral for Quasi-Hermitian Hamiltonians Represent?}
\author{H.~F.~Jones\footnote{e-mail: h.f.jones@imperial.ac.uk}\; and R.~J.~Rivers\footnote{e-mail: r.rivers@imperial.ac.uk}\\
Physics Department, Imperial College, London SW7 2AZ, UK}
\date{\today}
\maketitle
\begin{abstract}
In the context of quasi-Hermitian theories, which are non-Hermitian in the conventional sense,
but can be made Hermitian by the introduction of a dynamically-determined metric $\eta$, we address the problem of how the
functional integral and the Feynman diagrams deduced therefrom ``know" about the metric. Our investigation is triggered by
a result of Bender, Chen and Milton, who calculated perturbatively the one-point function $G_1$ for the quantum Hamiltonian
$H=\half(p^2+x^2)+igx^3$. It turns out that this calculation indeed corresponds to an expectation value in the ground state
evaluated with the $\eta$ metric. The resolution of the problem turns out be that, although there is no explicit mention of
the metric in the path integral or Feynman diagrams, their derivation is based fundamentally on the Heisenberg equations of
motion, which only take their standard form when matrix elements are evaluated with the inclusion of $\eta$.
 \\

%\noindent \footnotesize{Keywords: Hamiltonian, quasi-Hermitian, metric, path integral}
\end{abstract}

\section{Introduction}
\setcounter{footnote}{1}
The original paper of Bender and Boettcher\cite{BB} gave convincing numerical and analytic evidence
that a class of non-Hermitian Hamiltonians can nonetheless possess a completely real spectrum, in their case
because of an unbroken $PT$ symmetry. However, at first sight such theories, treated as fundamental quantum
mechanical theories, are still not acceptable because the natural Hilbert-space metric, whereby the overlap
between two wave functions $\chi$ and $\psi$ is not the standard $\int dx \chi^*(x)\psi(x)$, but rather
$\int dx \chi^*(-x)\psi(x)\equiv\int dx (\chi)_{PT}\psi$,  is non-positive-definite and so does not
lend itself to the usual probabilistic interpretation. Such a probabilistic interpretation was subsequently
resurrected by the realization\cite{BBJ1,AM1} that a positive-definite metric could in fact be constructed,
although it is not determined once and for all, like the standard metric, but depends on the Hamiltonian itself, and therefore differs from one model to another.

The authors of Ref.~\cite{BBJ1} introduced the so-called $C$ operator, with coordinate-space representation $C(x,y) = \sum_n\phi_n(x)\phi_n(y)$ in terms of the $PT$-invariant eigenstates $\phi_n(x)$ of $H$, and the overlap
$\int dx (\chi)_{CPT}(x)\psi (x)$, where $(\chi)_{CPT}(x) = \int dy\, C(x,y)\chi^*(y)$.
It was subsequently\cite{BBJ2} found useful to write $C$ in exponential form, $C=e^{Q}P$. The $Q$ operator provides
the link with the notation of Mostafazadeh\footnote{Strictly speaking we should write $\eta_+$ for $\eta$, to emphasize
the fact that it is positive-definite}, namely $\eta=e^{-Q}$. In this notation, the matrix element of any operator
${\cal O}$ is
\bea\label{db}
\langle\langle \chi|{\cal O}|\psi\rangle\rangle=\langle\chi|\eta{\cal O}|\psi\rangle.
\eea
$C$ commutes with $H$, which translates into the statement of quasi-Hermiticity:
\bea\label{qH}
H^\dag=e^{-Q} H e^Q\equiv \eta H \eta^{-1}.
\eea
As realized by Mostafazadeh, this latter equation can be used to define an equivalent Hermitian Hamiltonian $h$,
with the same spectrum as $H$, by the similarity transformation
\bea\label{sim}
h=e^{-\half Q} H e^{\half Q}\equiv \rho H \rho^{-1}.
\eea
The terminology ``quasi-Hermitian" is now commonly used to describe non-Hermitian Hamiltonians $H$ thus related to a Hermitian Hamiltonian $h$.
The metric $\eta$ is central to the formulation of such theories in the Schr\"odinger formalism.
However, its role in the functional integral formalism is much less prominent, even subliminal\cite{JR,AM2}.
Nonetheless the functional integral, and the Feynman rules derived from it, do somehow ``know" about the metric.

The present investigation is prompted by the calculation by Bender et al.\cite{BCM} of the one-point function for
the imaginary cubic Hamiltonian,
\bea
H=\half(p^2+x^2) +igx^3,
\eea
corresponding to the Lagrangian
\bea
L=\half(\dot{x}^2-x^2)-igx^3,
\eea
using the
standard Feynman rules derived therefrom. The result, to order
$g^3$ is
\bea\label{G1}
G_1=-\frac{3}{2}ig+\frac{33}{2} i g^3\ .
\eea
It turns out that this is the result of calculating the Green function, using the formula of Eq.~(\ref{db}),
as the expectation value of ${\cal O}=x$ in the ground state $|\chi\rangle=|\psi\rangle=|\psi_0\rangle$. In other words,
the graphical calculation of $G_1$ corresponds specifically to the matrix element calculated using the $\eta$
metric, even though that metric appears nowhere in the former calculation.

It is the purpose of the present paper to elucidate this problem, and answer the question posed in the title.
The plan of the paper is as follows: in the next section we look at a simpler problem, the Swanson model\cite{MS}, where
we do not even have to deal with perturbation theory. In Section 3 we discuss in more general terms the relations
between the Schwinger-Dyson equations and perturbation theory for the imaginary cubic potential. In
Section 4 we see how the same results arise from a functional integral. In Section 5 we summarize what we have learned from this analysis. The Appendix deals with the relation between the Schr\"odinger and Heisenberg pictures for quasi-Hermitian theories, which turns out to be crucial.

\section{The Swanson Model}
The classical Hamiltonian for this model is
\bea
H=a x^2+b p^2+ 2c x p,
\eea
where $a$ and $b$ are real and positive, while $c$ is pure imaginary.
The construction of $Q$ is not unique.

\subsection{$Q=Q(x)$}\label{Qx}
 We first suppose that $Q=Q(x)$.
The associated Lagrangian is obtained by setting $\dot{x}=\partial H/\partial p=2(b p+c x)$, to give
\bea
L=p\ \dot{x}-H = \frac{\dot{x}^2}{4b}-\tilde{a} x^2,
\eea
where $\tilde{a}=a-c^2/b$, after having discarded a total time derivative. This latter is in fact a multiple of the
derivative of the $Q$ operator, when this is taken to be a function of $x$ only\cite{JR}. Thus the Lagrangian $L$
associated with $H$ is the standard Hermitian Lagrangian of a scaled harmonic oscillator. Hence both Feynman diagrams
and Green functions may be evaluated within the standard framework of quantum theory, for which the Hermitian Hamiltonian
$h$ corresponding to $L$ is (with $P=\partial L/\partial \dot{x}=\dot{x}/(2b)=p+(c/b) x$)
\bea
h(x,P)=b P^2+\tilde{a}x^2.
\eea
Therefore the Green functions $G_n(t_1, \dots t_n)$ are just standard vacuum expectation values of time-ordered products
in the vacuum state. For simplicity let us just consider $G_2(t,t)$, which is
\bea
G_2(t,t)=\langle \Omega_h|x^2|\Omega_h\rangle
\eea
The ground state of the original non-Hermitian $H$ is related to that of $h$ by the similarity transformation of Eq.~(\ref{sim}),
namely
\bea
|\Omega_h\rangle=e^{-\half Q}|\Omega_H\rangle\ .
\eea
Thus
\bea
G_2(t,t)=\langle \Omega_H|e^{-\half Q}x^2e^{-\half Q}|\Omega_H\rangle
\eea
In this case, because $Q=Q(x)$ it commutes with $x$, so that
\bea
G_2(t,t)=\langle \Omega_H|e^{-Q}x^2|\Omega_H\rangle= \langle\langle\Omega_H| x^2|\Omega_H\rangle\rangle
\eea
Thus the Green function associated with the Lagrangian $L$ is unequivocally the vacuum expectation
value of $x$ evaluated using the $\eta$ metric.

\subsection{$Q=Q(p)$}
 Although $Q=Q(x)$ was the simplest possibility,  there is a one-parameter family of $Q$ operators\cite{HFJ, MZ},
where $Q$ can depend on both $x$ and $p$. Perhaps the extreme opposite of the previous case is when $Q=Q(p)$. As we have seen,
a straightforward calculation of the Lagrangian leads us to $Q=Q(x)$, so how can we access the other cases? The answer is that
we must first make a change of variables.

For the case $Q=Q(p)$ we consider the Hamiltonian $H(\xi,p)$, with $\xi$ yet to be specified, and make the shift
\bea\label{shift}
\xi=x-(c/a)p\ .
\eea
Then $H$ becomes
\bea
H(\xi,p)=ax^2+\tilde{b}p^2\equiv h(x,p),
\eea
where $\tilde{b}=b-c^2/a$, with corresponding Lagrangian
\bea
\ell(x)=\frac{\dot{x}^2}{4\tilde{b}}-a x^2
\eea
Now we have a new Hermitian harmonic oscillator, which, however, is a scaled version of the previous one,
since $a\tilde{b}=\tilde{a}b$. Thus the ground states $|\Omega_h\rangle$ are identical.

Proceeding as in \S\ref{Qx} to derive Green functions derived from the functional integral or Feynman diagrams
for $H(\xi,p)$, we will obtain standard expectation values of products of $\xi(t_r)$ with respect to the ground
state $|\Omega_h\rangle$, so in particular
\bea
G_2(t,t)&=&\langle \Omega_h|\xi^2|\Omega_h\rangle\nonumber\\
&=&\langle \Omega_H|e^{-\half Q}
\xi^2e^{-\half Q}|\Omega_H\rangle.
\eea
At this point it is useful to supplement Eqs.~(\ref{qH}) and (\ref{sim}) with the more general statements that an observable $A$, with real eigenvalues, is quasi-Hermitian:
\bea
A^\dag=e^{-Q} A e^Q
\eea
and is related to its Hermitian counterpart $a$ by
\bea
A&=&e^{Q/2}a e^{-Q/2},
\eea
or equivalently
\bea\label{dag}
A^\dag&=&e^{-Q/2}a e^{Q/2}.
\eea
In the present case $Q(p)$ is actually $Q=i(c/a)p^2$, which gives $X=x+(c/a)p$ for the observable associated with $x$.
Thus we see that in Eq.~(\ref{shift}) the variable $\xi$ must be identified with $X^\dag$. Accordingly
\bea
G_2(t,t)&=&\langle \Omega_H|e^{-Q}\left(e^{\half Q}(X^\dag)^2e^{-\half Q}\right)|\Omega_H\rangle\nonumber\\
&=&\langle \Omega_H|e^{-Q} x^2|\Omega_H\rangle=\langle\langle \Omega_H|x^2|\Omega_H\rangle\rangle \ .
\eea
What we have effectively done is to make use of the second of the identities
\bea\label{identities}
H(x,p)&=&e^{-Q/2}h(x,p)e^{Q/2}=h(X,P),\nonumber\\ && \\
H(X^\dag,P^\dag)&=&e^{-Q/2}h(X^\dag,P^\dag)e^{Q/2}=h(x,p),\nonumber
\eea
where, in this particular case, $P\equiv p$. A completely parallel treatment of case \ref{Qx} would
have started with $H(x,\pi)$, where $\pi=P^\dag$. However, we did not do this because the momenta drop out of the calculation
immediately $L$ is formed. With this understanding of the $\eta$ metric in mind, we return to the problem as posed, the Feynman diagram representation of Green functions.

\section{The Schwinger-Dyson Equations}

We consider a general non-Hermitian Hamiltonian of the form
\bea
H=\half(p^2+x^2)+ U(x).
\eea
Let us assume for the moment that the Green functions are calculated using the $\eta$ metric, and verify later that that is
indeed the case. Then the vacuum-generating functional is
\bea\label{Zdef}
Z[j] &=& \langle \Omega| \eta T(\exp[i\int dt j(t) x(t)])|\Omega\rangle,
\eea
which by successive functional differentiation with respect to the external current $j(t)$ gives
\bea
 G_n(t_1t_2...t_n) &\equiv& \langle \Omega|\eta T(x(t_1) x(t_2)...x(t_n))|\Omega\rangle\nonumber\\ &&\\
 &=& \frac{1}{Z[j]}(-i\delta /\delta j(t_1 ))(-i\delta /\delta j(t_2 ))...(-i\delta /\delta j(t_n )) Z[j]\bigg |_{j =0}\nonumber
\eea
For a given $U$ the Schwinger-Dyson equations are derived from the operator-valued equation of motion (see App.~A)
\bea\label{xeq}
(\partial_t^2+ 1)x(t)+ U'(x(t)) = 0.
\eea

Since $\eta$ is time independent this gives rise to a functional differential equation for $Z[j]$, namely
\bea\label{Zeq}
\left[(\partial_t^2 + 1)(-i\delta /\delta j(t )) + U'(-i\delta /\delta j(t))\right]Z[j] = j(t)Z[j].
\eea
We stress that this last equation contains no reference to $\eta$ and is just as we would expect for a standard Hermitian theory.
However, had we not included the correct metric $\eta$ we would not have obtained the usual
Heisenberg equations of motion.

For the case of interest, $U=igx^3$, we can obtain the Schwinger-Dyson equations by repeated functional differentiation of  (\ref{Zeq}) with respect to $j(.)$ at $j = 0$.
In terms of the propagator $\Delta(t-t')$, the solution to $(\partial_t^2 +1)\Delta(t -t') = -\delta (t - t')$
with appropriate boundary conditions, the first few are:
\bea\label{SD1}
G_1(t) &=& 3ig\int dt'\Delta(t-t')G_2(t't') = 0, \nonumber
\\
G_2(tt_1)&=& i\Delta(t-t_1) +  3ig\int dt'\Delta (t-t')
G_3(t't't_1),
\\
G_3(tt_1t_2) ) &=&
i[\Delta(t-t_1)G_1(t_2)+(t_1\leftrightarrow t_2)] + 3ig\int
dt'\Delta(t-t') G_4(t't't_1t_2), \nonumber
\\
&&\dots\dots\nonumber
\eea
Explicitly, $\Delta(t)=-\half i e^{-i|t|}$.
Note that $G_1(t)=\langle\langle x(t)\rangle\rangle\equiv\bar{x}$ is in fact a constant by
time-translation invariance.\\
\\
Here the $G_n$ are not connected. The relations to the connected Green functions $W_n$, with generating functional
\bea
W[j] = -i \ln Z[j],
\eea
are,
with $G_1 = W_1 = {\bar x}$, and
 $W_2(tt')\equiv W_2(t-t')$,
\bea\label{SDi}
& &{\bar x} = 3ig(\int dt'\Delta(t-t'))[{\bar x}^2 + W_2(0)] = -3ig[{\bar x} ^2 + W_2(0)]\nonumber,
\\
& &W_2(t-t_1)  = i\Delta(t-t_1) +  3ig\int dt'\Delta(t-t')
[2{\bar x}W_2(t' - t_1) + W_3(t't't_1)],
\\
& &W_3(tt_1t_2) ) =
   3ig\int dt'\Delta(t-t')[2{\bar x}W_3(t't_1t_2) + 2 W_2(t't_1)W_2(t't_2) + W_4(t't't_1t_2)]\nonumber,
\\
& &\dots\dots
\nonumber
\label{SD2}
\eea
 In the final equation we expanded about the time $t$. All these, and further equations, have simple diagrammatic
 representations.
 We stress that these are the conventional combinatoric equations for connected Green functions in an $x^3$ theory with
 imaginary coupling constant.
\\
\\
The Schwinger-Dyson equations (\ref{SDi}) do not by themselves give a unique solution. Thus, a priori $W_1$
is completely undetermined, and $W_2$ is only partially. However, they have a unique
perturbative expansion in $g$ \cite{RCUP}.
The lowest terms in standard perturbation theory follow from the first and second of Eqs.~(\ref{SDi}),
namely
\bea
W_2(t)&=&i\Delta(t)+O(g^2)\nonumber\\
\bar{x}&=&-3igW_2(0)+O(g^3)=-\frac{3}{2}ig+O(g^3),
\eea
the latter corresponding to the tadpole diagram of Ref.~\cite{BCM}.

As mentioned in the Introduction, and as already remarked in Ref.~\cite{HFJ}, this value unambiguously corresponds in the
Schr\"odinger picture to the expectation value
\bea
\bar{x}&=&\langle\psi_0|e^{-Q}x|\psi_0\rangle\nonumber\\
&=&\int dx\ \psi^*_0(x)(1-gQ_1)x\psi_0(x),
\eea
where
\bea
Q_1=-\frac{4}{3}p^3-2xpx
\eea
and $\psi_0(x)=e^{-\half x^2}/\pi^{\frac{1}{4}}$. Its non-zero value is entirely due to the presence of the $e^{-Q}$ factor.
We have also checked the $0(g^3)$ result reported in Eq.~(\ref{G1}). In that case we need both $Q_3$, occurring in the expansion $Q=gQ_1+g^3Q_3+\dots$, and the third-order
wave function.

We can now identify the reason why the perturbation expansion of the Schwinger-Dyson equations and the resulting Feynman
diagrams correspond to matrix elements calculated using the $\eta$ metric. As detailed in Appendix A, it is only when matrix
elements are so calculated that the Heisenberg-picture fields obey the standard equations of motion, which are the starting
point for the Schwinger-Dyson equations and, subsequently, the functional-integral representation of Green functions and
 the Feynman diagram
expansion.

Note that as an expectation value in the ground state $|\vf_0\rangle$ of the equivalent Hermitian Hamiltonian
\bea
\bar{x}&=&\langle\psi_0|e^{-Q}x|\psi_0\rangle\nonumber\\
&=&\langle\vf_0|e^{\half Q} e^{-Q}xe^{\half Q}|\vf_0\rangle\nonumber\\
&=&\langle\vf_0|X^\dag|\vf_0\rangle\
\eea
This latter calculation was also performed diagrammatically in Ref.~\cite{BCM}. Their variable $\tilde{x}$ is in fact
$X^\dag$.

In the present case there is another, non-perturbative, solution to Eqs.~(\ref{SDi}) starting with $\bar{x}=i/(3g)$, which is actually the
lower minimum of the classical potential. However, after making the shift
\bea
x=\frac{i}{3g}+y
\eea
the Hamiltonian becomes
\bea
H=\half(p^2-y^2)+igy^3+\ const.,
\eea
which corresponds to a theory with the wrong sign of the quadratic term and is hence unacceptable.

\section{The Functional Integral}

The formal solution to Eq.~(\ref{Zeq}) in terms of a path integral is
\bea\label{PIx}
Z[j] = \int [D x]\,\exp\left\{i\int dt \left[\frac{1}{2}{\dot x}^2 - \frac{1}{2}x^2 - U(x) + jx\right]  \right\}
\eea

In general one might have to worry about possible contours in the complex $x$ plane on which the functional integral is defined.
However, this is not an issue if we are restricting our attention to perturbation theory, since by Eq.~(\ref{PIx}) we mean the perturbative expansion
\bea
Z[j] = \int [Dx]\,\sum_n \frac{(-U(x))^n} {n!}\ \exp\left\{i\int dt \left[\frac{1}{2}{\dot x}^2 - \frac{1}{2}x^2 + jx\right]
\right\},
\label{pert}
\eea
with each term in the series a derivative of the well-defined
Gaussian integral
\bea\label{PIx0}
Z_0[j] &=& \int [Dx]\,\exp\left\{i\int dt \bigg[\frac{1}{2}{\dot x}^2 - \frac{1}{2}x^2 + jx\bigg]
\right\}\nonumber\\
&=&\exp\left[-\half i\int dt dt' j(t')\Delta(t-t')j(t)\right],
\eea
Although (\ref{pert}) recreates the perturbation series for the Schwinger-Dyson equations  trivially, there is still no guarantee that such a divergent series is asymptotic. For the case in point $U(x) = igx^3$ we know that the metastability of the corresponding real potential $U(x) = g x^3$ enforces a lack of Borel summability \cite{McKane} but we expect the complex potential to give Borel resummable series.

The above concerns expectation values of products of $x(t_i)$. However, suppose we wished to do the same for other variables, in
particular the corresponding observable $X$, which typically is a function of both $x$ and $p$. In that case we would have to start
with the equivalent Hermitian Hamiltonian $h(x,p)$, and exploit identities such as
\bea
\langle\langle\Omega_H|X^2|\Omega_H\rangle\rangle=\langle\Omega_h|x^2|\Omega_h\rangle
\eea
Because $h(x,p)$ typically contains higher powers of $p$, one needs to use the full phase-space functional integral
representation of $Z[j]$, namely
\bea
Z[j]=\int [Dx] [Dp] \exp\left\{-i\int dt\left[p\, \dot{x}+h(x,p)- j\, x\right]\right\},
\eea
subject to the usual ordering problems. To rewrite this in terms of $H$ one then needs to change variables to
$\xi=X^\dag$, $\pi=P^\dag$ using the identity $H(\xi,\pi)=h(x,p)$. This is what was done for the Swanson model in
Ref.~\cite{JR}. In Section 2 we were instead concerned with expectation values of products of $x$ and so started
with $\langle\Omega_h|\xi^2|\Omega_h\rangle$.

\section{Discussion}

Our main purpose in this paper was to understand how the perturbative calculation by Bender et al. \cite{BCM} picked out a
particular version of the one-point Green function for the non-Hermitian theory with $U=igx^3$,
namely $G_1=\langle\psi_0|\eta x |\psi_0\rangle$, rather than $\langle\psi_0|x |\psi_0\rangle$,
in spite of the fact that the metric $\eta\equiv e^{-Q}$ does not appear anywhere in the Feynman diagram expansion.

After looking at the simpler soluble problem of the Swanson Hamiltonian, where we verified that the same feature occurred
there for the two-point function $G_2(x,x)$, we went back to the derivation of the Schwinger-Dyson equations, which in turn
lead to the functional-integral representation of the vacuum generating functionals $Z[j]$ and $W[j]$, and to their perturbative
expansions.
Even then, the solution is not necessarily unique because of the problem of
giving meaning to a divergent series (see \cite{RCUP} for a more detailed discussion).

Starting from the Heisenberg equation of motion for the field $x(t)$, Eq.~(\ref{xeq}), it is in fact the case that one would
obtain the functional differential equation for $Z[j]$, Eq.~(\ref{Zeq}), whether or not the metric $\eta$ was included in the
definition of $Z[j]$ in Eq.~(\ref{Zdef}). The crucial point is rather that, as shown in the Appendix, these equations of motion
will only take their standard form $i\dot{A_H}=[A_H,H]$ in a quasi-Hermitian theory, if the relation between the Schr\"odinger
and Heisenberg pictures is built on matrix elements defined with the inclusion of the metric.

Whether or not these Green functions, being time-ordered expectation values of products of the non-observable field $x(t)$,
are useful is another question. It is worth remarking that in our previous paper\cite{JR} on functional integrals for
non-Hermitian Hamiltonians we were instead concerned with the functional integral representation of Green functions of
observables, such as $\langle\langle \Omega_H|X^2|\Omega_H\rangle\rangle$ in Case 2.2 for the Swanson Hamiltonian.

Finally, there is nothing peculiarly quantum mechanical, as distinct from quantum field theoretical, about our analysis, which extends naturally to pseudo-Hermitian quantum field theory.

\section*{Appendix}

{\bf Schr\"odinger and Heisenberg Pictures}\\

Let us denote a Schr\"odinger-picture state by $|\vf\rangle$ and its corresponding Heisenberg state by $|\psi\rangle$, so that
\bea
|\psi\rangle=e^{-iHt}|\vf\rangle,
\eea
identifying the two pictures at $t=0$.
Correspondingly
\bea
\langle\psi|=\langle \vf|e^{iH^\dag t}
\eea
We identify the Heisenberg operator $A_H(t)$ corresponding to a given Schr\"odinger operator $A$ by considering
\bea
\langle\psi|\eta A|\psi\rangle&=&\langle\vf| e^{iH^\dag t}\eta A e^{-iHt}|\vf\rangle\nonumber\\
&=&\langle\vf|\eta e^{i Ht}A e^{-iHt}|\vf\rangle,
\eea
by virtue of Eq.~(\ref{qH}). Thus we identify
\bea\label{Heisenberg}
A_H=e^{iHt}Ae^{-iHt},
\eea
the usual transformation. However, it is no longer unitary because $H$ is not Hermitian.
Since we have no explicit time dependence in the theory, $\eta$ itself is time independent because
$e^{iH^\dag t}\eta e^{-iHt}=\eta$.
From Eq.~(\ref{Heisenberg}) we obtain the Heisenberg equation of motion for $A_H$:
\bea
i\frac{d}{dt}A_H=[A_H, H].
\eea
This equation does not, however, apply to $\eta$.


\begin{thebibliography}{99}
\bibitem{BB} C.~M.~Bender and S.~Boettcher, Phys.~Rev.~Letter {\bf 80} (1998) 5243.
\bibitem{BBJ1} C.~M.~Bender, D.~C.~Brody and H.~F.~Jones, Phys.~Rev.~Letter {\bf 89} (2002) 270401;
{\bf 92} (2004) 119902(E).
\bibitem{AM1} A.~Mostafazadeh, J.~Math.~Phys. {\bf 43} (2002) 205;
J.~Phys. A {\bf 36} (2003) 7081.
\bibitem{BBJ2} C.~M.~Bender, D.~C.~Brody and H.~F.~Jones,
Phys.~Rev.~D {\bf 70} (2004) 025001; {\bf 71} (2005) 049901(E).
\bibitem{JR} H.~F.~Jones and R.~J.~Rivers, Phys.~Rev.~D {\bf 75} (2007) 025023.
\bibitem{AM2} A.~Mostafazadeh, Phys.~Rev.~D {\bf 76} (2007) 067701.
\bibitem{BCM} C.~M.~Bender, J.-H.~Chen and K.~A.~Milton, J.~Phys. A {\bf 39} (2006) 1657.
\bibitem{MS} M.~S.~Swanson, J.~Math.~Phys. {\bf 45} (2004) 585.
\bibitem{HFJ} H.~F.~Jones, J.~Phys. A {\bf 38} (2005) 1741.
\bibitem{MZ} D.~P.~Musumbu, H.~B.~Geyer and W.~D.~Heiss, J.~Phys. A {\bf 40} (2007) F75.
\bibitem{RCUP} R.~J.~Rivers,{\it Path Integral Methods of Quantum Field Theory}, CUP (1987),  Section 1.4.
\bibitem{McKane} A.~J.~McKane, Nucl. Phys. B {\bf 152} (1979) 166.
\end{thebibliography}
\end{document}